\documentclass[11pt]{article}
\usepackage[margin=1in]{geometry}
\usepackage[T1]{fontenc}
\usepackage[utf8]{inputenc}
\usepackage{microtype}
\usepackage{booktabs}
\usepackage{array}
\usepackage{tabularx}
\usepackage{flafter}
\usepackage{float}
\usepackage{enumitem}
\usepackage{hyperref}
\usepackage[round,authoryear]{natbib}
\hypersetup{colorlinks=true,allcolors=blue}
\setlist{nosep}

\title{When Agents Act Unwatched: The Reduced-Supervision Paradox in Agentic AI}
\author{Hanjing Shi \and Dominic DiFranzo\\
Department of Computer Science and Engineering\\
Lehigh University, Bethlehem, PA, USA}
\date{August 2026}

\begin{document}
\maketitle

\begin{abstract}
Agentic AI is sold on a simple promise: the system keeps acting when the user stops watching. That promise creates an accountability inversion. As stepwise supervision recedes, verification does not disappear; it moves into the runtime infrastructure that defines authority, records action, interrupts execution, checks outcomes, and supports repair. We call this the \emph{reduced-supervision paradox}. Using a 63-artifact audit, we examine its public visibility across 46 research papers and 17 engineering, documentation, security, and governance sources. We find that agents' action surfaces are far easier to reconstruct than the mechanisms needed to answer for their actions. Tool mediation and monitoring traces were clearly visible in 40 and 37 artifacts, whereas checkpoint placement was clearly visible in 6, validator independence in 4, recovery in 2, and contestability in 1. Three action paths show why this imbalance matters. A repository path can preserve rich diffs after a consequential change. A browser path can cross organizational boundaries faster than permissions travel. A service path can follow policy while leaving affected people without recourse. We argue that observability can become a substitute for accountability when it shifts verification onto users after meaningful intervention is no longer possible. Our action-path diagnostic instead asks whether a delegated action remains connected to authority, evidence, interruption, independent judgment, recovery, and challenge. The claim is deliberately bounded to public visibility; it does not establish the prevalence or effectiveness of undisclosed controls. We contribute an action-level account that recasts the harness from a technical wrapper into accountability infrastructure.
\end{abstract}


\section{Introduction}

An agent earns its name by continuing. A coding agent can inspect a repository, edit several files, run tests, and revise its work. A browser agent can navigate forms, retrieve records, and submit changes. A service agent can modify a booking. The user does not approve every token or tool call. That reduction in stepwise supervision is part of the product.

It also moves the accountability problem. When a person watches each step, they can interrupt before a consequential action. When the system continues alone, the person must rely on the surrounding runtime to preserve boundaries and evidence. Permissions define what the agent may touch. Traces record what it did. Checkpoints determine when a person or policy can stop it. Validators decide whether the work is acceptable. Recovery and contestation determine what happens after the wrong action reaches the world.

This is the reduced-supervision paradox. The practical value of agentic AI increases as immediate human inspection recedes, while accountability depends more heavily on the infrastructure that replaces that inspection. The user delegates action and, often, much of the process for knowing whether the action was justified. If the public record describes tool access and task completion but leaves checkpoints, independent validation, repair, and recourse vague, the system can become technically observable without becoming answerable.

The problem extends earlier work on automation reliance. Parasuraman and Riley show how automation can be misused or disused when people calibrate reliance poorly \citep{parasuraman1997humansandautomation}. Bu\c{c}inca and colleagues show that people may accept an AI recommendation when the effort of checking exceeds the effort of relying \citep{bucinca2021trust}. Agentic workflows intensify this older problem because the system does not merely recommend. It can change external state. A coding agent edits files. A browser agent submits forms. A service agent changes a booking.

Research on tool-using agents makes the action surface increasingly concrete. ReAct interleaves reasoning with external action \citep{yao2022react}. Toolformer incorporates API calls into model behavior \citep{schick2023toolformer}. Voyager shows how executable skills, environment feedback, and memory support extended embodied action \citep{wang2023voyager}. AutoGen makes orchestration programmable through conversations among agents, tools, and people \citep{wu2023autogen}. SWE-agent structures software work through an agent-computer interface for repository navigation, editing, and testing \citep{yang2024swe}. OpenHands extends that work into a general platform where agents write code, use a command line, and browse the web \citep{wang2024openhands}. OSWorld evaluates agents across real desktop and web applications \citep{xie2024osworld}. These projects do more than demonstrate capability. They show that capability is mediated: a model acts only through an environment that decides what it can see, which tools it can call, what state persists, and what evidence remains.

We call that runtime arrangement the \emph{harness}. The harness connects model output to external action through tool adapters, permissions, memory, orchestration, policy checks, traces, validators, handoffs, escalation, and recovery. The term is sometimes used narrowly for prompt and context management. Here it names the layer where delegated authority becomes executable and where an organization can still shape what happens before, during, and after action.

The harness matters because the delegating user is not always the only person affected. A repository change may alter a colleague's work. A booking agent may cancel travel for another passenger. A workplace agent may edit a shared record or send a message in an employee's name. An operator can have a complete technical trace while the affected person has no notice, action identifier, or correction route. Accountability therefore requires more than logging.

We study what public artifacts make visible about this layer. The corpus contains 63 artifacts: 46 research papers and 17 non-research sources. We ask whether authority, execution evidence, intervention, independent judgment, and post-action remedy remain reconstructable.

Three questions guide the analysis:

\begin{enumerate}[label=RQ\arabic*.]
\item Which harness functions are clearly visible in public artifacts about delegated-action systems?
\item How do those artifacts connect action evidence to checkpoints, independent validation, recovery, and contestation?
\item What does an action-path analysis reveal for users, operators, auditors, and affected third parties that a feature inventory misses?
\end{enumerate}

The paper makes three contributions. First, it establishes a bounded empirical asymmetry: across a 63-item public corpus and a source-type sensitivity analysis, evidence of action is markedly easier to inspect than evidence of intervention, independent judgment, repair, or challenge. Second, it develops the reduced-supervision paradox as an account of displaced verification. The paradox makes a stronger claim than a general warning about autonomy: observability can legitimize delegated action while leaving the burden of proving and repairing error with the people least able to inspect the process. Third, the paper contributes an action-path diagnostic that follows one delegated act through permission, action, trace, checkpoint, validation, recovery, and contestation. The diagnostic does not certify safety or predict incidents. It asks whether an action remains answerable from authorization to repair, for the user, operator, auditor, and affected third party who may each need different evidence.

\section{From Model Capability to Harnessed Action}

\subsection{Agents act through an environment}

An agent's apparent autonomy is built from concrete arrangements. The environment chooses the available tools, defines schemas, manages credentials, returns observations, retains memory, and decides whether a command executes. Agent capability is therefore mediated capability. The model proposes or selects an action; the harness makes the proposal legible to an external system and determines the conditions under which it changes state.

This mediation is visible across technical traditions. Tool-use research studies how models learn or select calls \citep{Schick_2023}. SWE-agent shows how an agent-computer interface structures repository navigation, shell commands, and test execution \citep{Yang_2024}. OpenHands places coding, command-line work, and browser interaction in a shared agent platform \citep{wang2024openhands}. ToolSandbox exposes stateful tool-use tasks and intermediate milestones \citep{lu2024toolsandbox}. AgentDojo tests how untrusted content can redirect tool-using agents \citep{debenedetti2024agentdojo}. InjecAgent benchmarks indirect prompt injection in tool-integrated agents \citep{Zhan_2024}. Tau-bench makes policy-constrained service workflows executable \citep{yao2024bench}. AutoGen extends mediation across interacting agents, tools, and human input \citep{wu2023autogen}. Bui shows how context and feedback scaffolding sustain terminal-based coding work \citep{bui2026building}. Symphony connects long-running coding tasks to issue trackers and isolated workspaces \citep{kotlyarsky2026open}.

The same mechanisms have governance effects. Tool schemas and credentials limit which actions can be requested and which resources can be changed, while sandboxes contain some side effects. Traces make later reconstruction possible. Checkpoints can keep risky proposals from executing, and snapshots can enable rollback. None of these properties belongs to the language model alone.

Model-centered evaluation remains necessary. It can measure planning, tool selection, instruction following, and task success. It can miss the arrangement in which the action is authorized, observed, and repaired. A model may perform well on a benchmark while a deployed harness grants credentials too broadly. It may produce an excellent plan while the checkpoint arrives after the irreversible step. It may generate a correct summary while the underlying trace is inaccessible to the user who needs to dispute the action.

The paper therefore treats the harness as an accountability object. This does not mean that accountability can be solved entirely in software. Organizations assign roles, set policies, and decide remedies. The harness is where those decisions become executable, recorded, or absent at the moment of action.

\subsection{Reduced supervision displaces verification}

Stepwise oversight is expensive. It interrupts long tasks, demands attention, and can convert automation into a stream of approval prompts. Agentic systems promise to reduce this burden. Coding agents work across context windows. Service agents interpret policies. Browser agents complete multi-page workflows. Their value often depends on not requiring the user to inspect every intermediate decision.

The verification labor does not vanish. It moves. Before execution, someone must determine the scope of authority and which actions need confirmation. During execution, the system must retain enough evidence to detect divergence and support intervention. After execution, someone must validate the result and repair unintended effects. During dispute, the organization must preserve evidence and offer a route to challenge.

This displacement creates two risks. The first is compression. A user receives a concise result while the system has taken many steps, touched multiple data sources, and crossed several policy boundaries. The summary can hide the point at which the decisive action occurred. The second is substitution. The same system that acted may also generate the explanation, validation, and confidence statement used to justify its action.

More supervision is not the automatic answer. Constant approval can produce habituation, fatigue, and superficial consent. The relevant question is checkpoint placement. Human or institutional attention should concentrate before consequential and difficult-to-reverse changes, at high uncertainty, or when the action affects people beyond the delegator. A checkpoint that appears after payment, message delivery, or record mutation records consent too late.

\subsection{Accountability requires an audience}

Accountability is procedural. An actor must be answerable to a forum that can question, judge, and require correction or another consequence \citep{bovens2007analysingandassessingaccountability}. A log may be evidence within that procedure. It is not the procedure itself.

Different audiences need different evidence. Developers need detailed trajectories, exceptions, and environment state. Operators need current risk signals and escalation status. Auditors need provenance, policy versions, permissions, and retained decisions. Users need a clear account of consequential actions and what can still be changed. Affected third parties may need only the action that involved them, the authority claimed, the responsible organization, and a route to correction.

One trace rarely serves all these audiences without translation. A full trajectory may expose sensitive data or proprietary reasoning. A compressed summary may omit the evidence needed to challenge a decision. Good reporting therefore requires role-specific views connected to the same action identifier and source record.

\subsection{From features to action paths}

A feature inventory asks whether the system has logging, guardrails, human oversight, or rollback. An action path asks whether those features remain connected for one consequential act. We define seven stages: permission, action, trace, checkpoint, validation, recovery, and contestation.

Permission identifies who authorized which scope. Action records the state-changing tool call or sequence. Trace preserves enough evidence to reconstruct the action. Checkpoint locates intervention before the relevant consequence. Validation checks the result with evidence sufficiently independent of the actor. Recovery stops, reverses, compensates for, or repairs the action. Contestation allows a user or affected party to challenge it and obtain review.

Continuity matters. A checkpoint can exist for file deletion but not message sending. A trace can record tool calls without the policy version that allowed them. A validator can inspect the final state but lack authority to stop deployment. Rollback can restore a repository while a sent message remains public. Contestability can exist in policy while the complainant lacks an action identifier. The path breaks at its weakest missing connection.

We use \emph{action surface} for tool access, execution, and traces that make what a system can do or did publicly visible. We use \emph{accountability closure} for checkpoints, independent validation, recovery, and contestation that allow an action to be questioned and repaired. Closure is an analytic standard, not a claim that every dispute has one final technical solution.

\subsection{Positioning against adjacent work}

Adjacent scholarship already supplies many of the pieces needed to analyze accountable agents, but it usually encounters the action at a different moment. Ezell and colleagues begin with an agent incident and specify the logs, system documentation, access, and tool information needed for investigation \citep{ezell2025incident}. Gailmard and colleagues ask how adverse-event reporting can scale when incidents and unknown failure modes accumulate \citep{Gailmard_2025}. Both approaches make post-incident evidence more systematic. Our action path begins earlier. It asks whether the authority, checkpoint, and independent judgment attached to an action can prevent or redirect the consequence before incident reporting becomes necessary.

Reviewability and traceability move the analysis beyond model outputs. Cobbe and colleagues treat automated decision-making as a sociotechnical process that begins before a decision and extends beyond it \citep{Cobbe_2021}. Kroll argues that traceability must connect how a system was built and operated to the governance purposes that make those records matter \citep{kroll2021outlining}. Chappidi and colleagues add a complication: record-keeping for accountability can reconfigure organizational work, create surveillance concerns, and generate resistance \citep{Chappidi_2025}. Kawakami and colleagues likewise show that responsible-AI artifacts do not automatically advance the goals of legal and civil-society stakeholders \citep{kawakami2024responsible}. These accounts establish that records are institutional interventions, not neutral containers. We extend that insight by asking whether the record for one delegated action remains connected to the permission that enabled it, the state change it produced, and the forum that can require repair.

Other work widens what this action-level account must include. Ibrahim and colleagues argue that static, model-only tests miss harms that emerge through sustained interaction \citep{Ibrahim_2025}. He and colleagues use least privilege to show why access should be limited to the information needed for a particular deployment \citep{He_2025}. Riedl and Desai argue that computer science accounts of agents under-theorize loyalty and relations with third parties \citep{riedl2025aiagentsandthelaw}. Alfrink and colleagues make contestability a lifecycle property rather than a post-hoc explanation feature \citep{Alfrink_2022}. We draw these concerns into one temporal unit. The issue is not only whether a system has interaction safeguards, scoped access, legal authority, or a challenge mechanism. It is whether those mechanisms remain attached to the same act as authority moves into execution and consequence.

Harness research comes closest to that temporal unit, but its primary audience is often the builder. Bui describes the scaffolding, context, and feedback loops that let coding agents persist across terminal work \citep{bui2026building}. Pan and colleagues treat natural-language harnesses as mechanisms for structuring agent behavior \citep{pan2026natural}. Symphony makes orchestration across long-running coding tasks explicit \citep{kotlyarsky2026open}. Control-plane proposals place policy enforcement at the point of tool execution \citep{developers2026securing}. Governance-by-architecture proposals similarly seek enforceable runtime constraints \citep{besanson2026sarc}. These approaches establish the harness as a substantive design object. We argue that a builder-facing control plane is still not an accountability path unless users, auditors, operators, and affected third parties can connect its controls to authority, evidence, and remedy.

This distinction also separates runtime continuity from document volume. AGORA makes a growing body of governance instruments searchable and comparable \citep{Arnold_2024}. Foundation-model transparency reports document provider practices \citep{Bommasani_2024}. Ecosystem graphs map dependencies across the model supply chain \citep{Bommasani_2024_2}. Platform-security analysis exposes risks distributed across models, plugins, and external services \citep{iqbal2024llmplatform}. Feffer and colleagues warn that even a widely endorsed practice such as red teaming can become security theater when its institutional role is underspecified \citep{Feffer_2024}. The action-path diagnostic narrows these infrastructure questions to a consequential test: can the same delegated action be followed across authorization, execution, evidence, intervention, and remedy, or does it disappear into separate technical logs, policy pages, support queues, and internal processes?

\subsection{Oversight, governance, and experience conditions}

The harness can be read through two runtime layers and one bounded downstream lens. The \emph{oversight layer} concerns when an agent may proceed, pause, ask, stop, or hand off. The \emph{governance layer} concerns who authorized the act, what record makes it reviewable, who may question it, and what repair remains possible. \emph{Experience conditions} concern what users and affected parties are positioned to understand or do when they encounter confirmations, status displays, explanations, handoffs, and correction routes. The corpus exposes oversight and governance mechanisms directly. It offers only indirect evidence about experience because it contains no user or affected-party data.

\paragraph{Oversight.} A human ``in the loop'' is not a meaningful control if the person receives the wrong evidence or arrives after the consequential state change. SABER makes this timing problem explicit by separating mutating actions from less consequential steps \citep{cuadron2025saber}. Tau-bench shows why the problem is also contextual: a service agent acts under domain policy while interpreting an underspecified user request \citep{yao2024bench}. Runtime oversight therefore requires an authority boundary, a risk-sensitive checkpoint, an independent basis for judgment, and a stopping effect. A read-only step may need no interruption. A state-changing step may require confirmation before execution.

\paragraph{Governance.} Oversight can stop an action; governance makes an organization answer for it. Authenticated-delegation work treats authority as scoped, auditable, and revocable rather than equating a user's broad intent with permission for every downstream act \citep{reference2025authenticateddelegationandauthorizedaiagents}. Our governance layer adds the record and forum that connect this authority to review and repair. A developer-facing trace may be technically complete yet inaccessible to an affected person. It may also be detached from the office, operator, or procedure empowered to provide a remedy.

\paragraph{Experience conditions.} A control that exists in the harness can still fail at the interface. Interactional-fairness work shows that procedural treatment matters within multi-agent encounters \citep{Binkyte_2025}. Practitioner-informed analysis of LLM counseling shows why domain obligations cannot be reduced to generic model behavior \citep{Iftikhar_2025}. Multilingual safety research shows that access to a system does not imply equal access to its safeguards \citep{Shen_2024}. Cognitive-forcing research shows that interruption helps only when it prompts meaningful engagement rather than another habitual click \citep{bucinca2021trust}. Public documentation cannot establish that people understand or use a control, but it can reveal whether the necessary conditions exist: visible scope, a timely confirmation, an intelligible status, and an accessible correction route.

The three layers make the argument cumulative. Oversight locates intervention before consequence. Governance connects that intervention and its evidence to an answerable organization. Experience conditions determine whether the people expected to rely on those arrangements can encounter and use them. Reduced supervision is accountable only when all three remain connected for the same delegated act.

\section{Method}

\subsection{Corpus and sampling boundary}

We analyzed 63 public artifacts released from 2020 onward. The corpus contains 46 research papers and 17 non-research artifacts. The latter include six technical-documentation sources, three engineering notes, three governance frameworks, one security taxonomy, one survey, one commentary, one policy commentary, and one related research article. The sample is purposive. It favors artifacts with enough runtime detail to inspect harness functions and is therefore visibility rich rather than representative.

Corpus construction began from seed sources on tool use, agent-computer interfaces, memory, execution environments, orchestration, security, and harness design. Searches covered Google Scholar, arXiv, ACM Digital Library, AI ethics and responsible-AI proceedings, ACL Anthology, OpenReview, Springer, official framework documentation, public engineering notes, institutional reports, and OpenAlex metadata. Query families included agent harness, harness engineering, orchestration, runtime governance, control plane, tool-execution policy, agent security, indirect prompt injection, incidents, authenticated delegation, contestable AI, adverse-event reporting, long-running agents, and runtime validation.

An artifact entered the corpus when it described a system that could act beyond single-turn text generation, exposed at least one runtime mediation mechanism, and contained enough detail to code at least two functions. Model-only benchmarks without a runtime layer, generic governance documents without an agent or deployment mechanism, interface descriptions with no connection to permission or evidence, and duplicate pages were excluded.

The final set spans coding agents, general agent frameworks, reasoning-and-acting systems, tool protocols, embodied agents, multi-agent systems, security and safety evaluation, incident analysis, law and delegation, runtime governance, enterprise web tasks, workplace agents, retail and airline service, computer use, scientific research agents, and governance frameworks. This heterogeneity is intentional. The study compares how different public genres expose the same runtime functions.

An OpenAlex coverage pass returned 1,044 query-level records and 917 records after deduplication. Fifteen high-salience candidates were read after metadata screening. They reinforced existing mechanism classes or fell outside the action-mediation boundary, so the final set remained 63. The coverage pass was not used as a prevalence denominator and did not change artifact codes.

\begin{table}[t]
\centering
\caption{Source composition of the 63-artifact corpus.}
\begin{tabular}{lr}
\toprule
Source class & Artifacts \\
\midrule
Research papers & 46 \\
Technical documentation & 6 \\
Engineering notes & 3 \\
Governance frameworks & 3 \\
Security, survey, commentary, policy, and related research & 5 \\
\midrule
Total & 63 \\
\bottomrule
\end{tabular}
\end{table}

\subsection{Coding fields and decision rules}

One primary coder recorded artifact type, domain, action scope, harness evidence, authority boundary, tool mediation, checkpoint placement, risk sensitivity, validator independence, monitoring and traces, escalation, recovery or rollback, contestability, third-party effects, privacy or surveillance, an evidence note, and an open gap. Governance fields used present, partial, unclear, absent, and not applicable.

\emph{Present} required explicit public evidence of the function and enough detail to recognize its role. \emph{Partial} marked a visible mechanism whose governance function remained underspecified. \emph{Unclear} meant that the artifact did not support a stronger inference. \emph{Absent} was reserved for settings in which the function applied but the artifact explicitly omitted it or made its absence visible. \emph{Not applicable} was used sparingly.

Tool mediation counted when the artifact identified tools, APIs, files, browsers, memory stores, command environments, or external systems through which the agent acted. Monitoring and traces required explicit histories, trajectories, event records, observability, or comparable evidence. A demonstration of final output alone did not count as a trace.

Checkpoint placement required a pause, verification, request, or review positioned before a consequential action. A general human-in-the-loop statement was partial unless timing and action class were clear. Risk sensitivity required a visible distinction among action types, uncertainty, stakes, or affected domains rather than a universal approval rule.

Validator independence required evidence that the checker was separated from the action-producing component or used a meaningfully different evidence source. Calling the same model again did not automatically count. The artifact had to identify the checker, evidence, separation, or stopping authority.

Recovery required a route to undo, compensate for, pause, or otherwise repair an external effect. Retries counted as reliability, not recovery, when the system simply attempted the same task again. Contestability required a route through which a person could challenge an action and obtain review or remedy. A feedback form without a documented review route was partial or unclear.

Boundary examples anchored these decisions. SABER counted as a checkpoint case because it describes mutation-gated verification before state-changing steps \citep{cuadron2025saber}. Tau-bench made policy-constrained service actions visible but did not itself document a deployed user approval or dispute workflow \citep{yao2024bench}. ReAct exposed action traces while leaving authority, recovery, and contestability outside its contribution \citep{yao2022react}.

The audit retains row-level sources, codes, evidence notes, and open gaps. This traceability supports reinspection. It does not establish intercoder reliability. No completed independent second-coder estimate is reported.

\subsection{Retrieval probe and source-type sensitivity}

A codebook-derived lexicon and BM25-style retrieval probe tested whether the same visibility pattern appeared in fetched public text or abstract snippets. Action-surface terms covered tools, calls, traces, trajectories, observation, and execution. Accountability-closure terms covered checkpoints, independent validation, rollback, compensation, appeal, contestation, and related language.

The probe did not assign codes or resolve ambiguous cases. It returned a mean hit rate of .49 for action-surface language and .11 for accountability-closure language. We use this result as a textual robustness check. Lexical retrieval cannot determine whether a checkpoint is well placed or a recovery mechanism is usable.

We also separated research and non-research artifacts. Among 46 research papers, tool mediation and traces were clearly present in 29 and 28, checkpoints in 5, validator independence in 3, recovery in 2, and contestability in none. Among 17 non-research artifacts, the corresponding counts were 11, 9, 1, 1, 0, and 1. Both source groups therefore showed the same direction despite different purposes and vocabularies.

\section{Findings}

\subsection{Action surfaces dominate the public record}

Tool mediation was clearly present in 40 of 63 artifacts, partial in 15, and unclear in 8. Monitoring or traces were clearly present in 37, partial in 23, and unclear in 3. These two fields made agent action concrete. Public readers could often identify the available tools, environment, trajectory, test loop, or event history.

The result reflects the technical center of agent research. A paper must explain what an agent can do and how task success is measured. Coding-agent artifacts show repository navigation, edits, tests, and pull-request workflows. Browser benchmarks show websites, forms, and state. Security benchmarks show attacker content, tools, and consequences. Engineering notes show observability, progress files, and orchestration.

These descriptions are valuable. They support reproduction, debugging, and evaluation. They also establish only the first half of accountability. Knowing that an agent called an API does not show who authorized the call, whether a checkpoint preceded it, whether the result was independently checked, or what happens if another person disputes the effect.

\subsection{Accountability closure is difficult to reconstruct}

Checkpoint placement was clearly present in 6 artifacts, partial in 40, unclear in 12, absent in 4, and not applicable in 1. Validator independence was clearly present in 4, partial in 36, unclear in 20, absent in 2, and not applicable in 1. Recovery or rollback was clearly present in 2, partial in 38, unclear in 20, absent in 2, and not applicable in 1. Contestability was clearly present in 1, partial in 28, unclear in 31, and absent in 3.

The large partial category is informative. Public artifacts frequently mentioned human review, guardrails, validation, rollback, or feedback. They less often specified when the review occurred, what evidence the validator used, which effects could be undone, or whether an affected person could obtain a reasoned response. Accountability language was present without procedural closure.

This pattern should not be read as a prevalence estimate. The underlying systems may contain undisclosed controls. Some artifacts were research papers whose contribution lay elsewhere. The result concerns what the public artifact allows a reader to reconstruct. On that measure, action is much clearer than repair.

\subsection{Authority and risk appear without a complete stopping rule}

Authority boundaries were clearly present in 17 artifacts, partial in 38, and unclear in 8. Risk sensitivity was more visible: 30 artifacts clearly distinguished stakes, action types, uncertainty, or dangerous conditions; 28 did so partially; 4 were unclear; and 1 was not applicable. These fields show that the corpus did not ignore permission and risk. The missing connection was procedural.

An artifact might state that an agent has limited tool permissions without explaining how a natural-language request maps onto those limits. It might classify a financial or mutating action as high risk without identifying the checkpoint that follows. It might require user approval while leaving the user to inspect a summary that omits affected accounts, downstream actions, or uncertainty. Public readers could often see that a boundary existed but could not reconstruct when it became binding.

Escalation showed the same pattern. Seven artifacts clearly described escalation, 43 described it partially, 10 were unclear, and 3 made its absence visible. A handoff is meaningful only if the trigger, recipient, evidence, and stopping effect are known. Escalation after an irreversible action can support incident response, but it does not substitute for a checkpoint. A handoff to a person who sees only the actor's summary may also preserve the same evidentiary blind spot.

Third-party effects were clearly present in 30 artifacts, partial in 15, and unclear in 18. The visibility of these effects did not usually produce a third-party procedure. An artifact could acknowledge that an agent touches customers, coworkers, or external services while keeping notification and contestation outside the runtime account. This gap explains why delegator-centered control is incomplete. The person who gives permission and the person who needs a remedy may be different.

Privacy and surveillance concerns were clearly present in 9 artifacts, partial in 38, and unclear in 16. This distribution complicates any proposal for richer traces. Public artifacts need to show not only what is recorded, but also why, for whom, and for how long. Otherwise, an effort to improve accountability can widen monitoring without improving a user's or affected party's ability to obtain correction.

These intermediate fields locate the break. Public artifacts frequently recognize authority, risk, escalation, third-party effects, and privacy. They less often connect those concerns to a timed checkpoint, independent evidence, repair, and a challenge route. The reduced-supervision paradox is therefore not a contrast between technical work and no governance language. It is a failure of continuity between recognized concerns and executable accountability.

\begin{table}[t]
\centering
\caption{Visibility of central harness functions in 63 public artifacts.}
\begin{tabular}{lrrrr}
\toprule
Function & Present & Partial & Unclear & Absent \\
\midrule
Tool mediation & 40 & 15 & 8 & 0 \\
Monitoring or traces & 37 & 23 & 3 & 0 \\
Checkpoint placement & 6 & 40 & 12 & 4 \\
Validator independence & 4 & 36 & 20 & 2 \\
Recovery or rollback & 2 & 38 & 20 & 2 \\
Contestability & 1 & 28 & 31 & 3 \\
\bottomrule
\end{tabular}
\end{table}

\subsection{Traces serve debugging before answerability}

Many artifacts described trajectories, test logs, tool histories, workflow events, repository state, or progress records. These traces help developers reproduce failure and help evaluators score tasks. They can be technically complete for those purposes while remaining institutionally incomplete.

A tool history may show that a message was sent. It may not show which user request authorized the recipient, which organizational policy permitted the action, whether private data was included, or who can retract the message. A repository trace may preserve every edit but not identify why a shared file fell within scope. A booking log may record an API call without presenting a traveler with the final fare or cancellation terms before purchase.

The gap is not solved by retaining everything. Rich logs can create privacy and surveillance risks. Worker actions, user content, credentials, and private records may become available to developers or auditors who do not need them. Accountability requires purpose limitation, role-based access, retention rules, and a way to extract the minimum evidence needed for a dispute.

The corpus makes this audience gap concrete. Several artifacts expose technically detailed trajectories without an equivalent record for the person whose work, account, or transaction changed. The design problem is therefore not simply how much to retain, but how to derive purpose-limited views from a common action record. Each view should remain tied to the same action identifier so that a later explanation cannot drift away from the underlying event.

\subsection{Validation often lacks visible independence}

The corpus frequently described testing, review, guardrails, critics, judges, or evaluators. Only four artifacts clearly exposed validator independence. The problem was not the absence of a check word. It was the absence of separation.

A validator can share the actor's blind spots when it uses the same model, context, retrieved evidence, and objective. It can also lack stopping authority. A post-hoc score may tell researchers whether a task succeeded while doing nothing to prevent a harmful mutation. An internal critic may improve an answer without checking whether the action was authorized.

Independence is contextual. A different model may still rely on the same incomplete record. A deterministic test can be independent of the model but irrelevant to social authority. A human reviewer can bring different judgment yet receive a compressed summary produced by the actor. Public documentation should therefore state four things: who or what validates, which evidence it receives, how that evidence differs from the actor's basis, and what happens when the check fails.

This requirement does not demand a human for every action. Low-stakes, reversible actions may use automated checks. Consequential mutations may require stronger separation. The point is visible allocation of verification, not universal manual approval.

\subsection{Recovery and contestation are weakest at the social boundary}

Technical artifacts often described retry, replanning, or error handling. These functions improve task completion. They do not necessarily repair external effects. Once an email is sent, a record overwritten, a refund issued, or a booking canceled, repeating the task may deepen the problem.

Only two artifacts clearly described recovery or rollback under the coding rule. Partial mechanisms included repository snapshots, sandbox resets, retries, and some escalation or handoff. Their scope was often local. A snapshot can restore files while leaving an external notification or downstream deployment unchanged.

Contestability was even less visible. The one clearly present case came from general contestable-AI design rather than a runtime agent artifact \citep{Alfrink_2022}. This location exposes a disciplinary split. Runtime work describes how the system acts. Governance work describes how decisions should be challenged. Few artifacts connect the challenge to the same action record, evidence, and repair mechanism.

The gap is most severe for people who did not delegate the agent. Tau-bench makes actions over orders and airline bookings visible \citep{yao2024bench}. WorkArena exposes changes made through enterprise web applications \citep{drouin2024workarena}. TheAgentCompany places agents in consequential workplace tasks involving shared systems and other organizational actors \citep{xu2024theagentcompany}. The affected person, however, may have no access to the agent interface or internal trace. Without notice and an external action identifier, they cannot even locate the process to contest.

\section{Three Action Paths}

The audit counts show a visibility asymmetry. Action paths show how it unfolds. The following examples synthesize public mechanisms already in the corpus. They are not incident case studies and do not claim that a particular deployed system followed every step.

\subsection{Repository change: a clean trace can end at merge}

Coding-agent artifacts make action highly inspectable. SWE-agent describes repository navigation, edits, shell commands, tests, and task outcomes \citep{yang2024swe}. OpenHands exposes similar actions through a general software-agent platform \citep{wang2024openhands}. Terminal-agent engineering adds planning, context management, and validation loops \citep{bui2026building}. Symphony connects issue trackers to isolated workspaces, progress artifacts, and human review \citep{kotlyarsky2026open}.

The path begins with permission. An issue or prompt authorizes work on a repository, but the exact scope may remain implicit. Does the agent have authority to modify generated files, dependencies, infrastructure, or security settings? Repository credentials answer what is technically possible. They do not explain what the task permits.

Action and trace are comparatively strong. File edits, commands, tests, and diffs can be retained. A developer can reconstruct how the patch changed. Checkpoint placement is less certain. Review often occurs at a pull request, which is useful when merge is the consequential action. It is too late if tests called external services, a command exposed a secret, or the agent already changed a shared branch.

Validation can also collapse into the action loop. The agent runs the tests that the repository provides and interprets the failures. This is valuable evidence, yet the actor and checker share the task environment. Independent validation may require protected tests, separate policy checks, static analysis, security review, or a human who sees the actual diff rather than the agent's summary.

Recovery is technically plausible because version control can revert a merge or restore a snapshot. The social effects may persist. Reviewers spend time, downstream automation runs, and a public release may be copied. Contestability remains least developed. A colleague whose file was changed can comment on a pull request, but public artifacts rarely specify rights, response times, or remedy for people affected outside the review interface.

The repository path is therefore not a failure of traceability. It is a case in which excellent traceability can stop short of authority and remedy. A complete public action record would connect the issue, permission scope, changed resources, external side effects, validation basis, merge checkpoint, recovery plan, and responsible reviewer.

\subsection{Enterprise browser work: the action crosses systems}

WorkArena evaluates agents on common knowledge-work tasks performed through enterprise web applications \citep{drouin2024workarena}. WorkArena++ adds compositional planning across those tasks \citep{boisvert2024workarena}. OSWorld extends the setting to workflows spanning real desktop and web applications \citep{xie2024osworld}. TheAgentCompany places agents inside a simulated software organization with consequential work \citep{xu2024theagentcompany}. AgentDojo adds untrusted content and indirect prompt injection to tool-mediated tasks \citep{debenedetti2024agentdojo}. These environments reveal a harder action path because state is distributed across systems.

Permission begins with account access. A user may authorize the agent to complete a task in one application. Single sign-on, stored sessions, and browser credentials can expose additional systems. The harness must translate a natural-language request into resource and action scopes. ``Update the project'' does not by itself authorize editing a coworker's record, reading an unrelated message, or submitting a form under another role.

The agent then navigates pages, retrieves context, enters data, and triggers actions. Traces can preserve clicks, page states, tool calls, and screenshots. They may include sensitive content and still omit the institutional meaning of a field. A browser knows that a button says submit. It may not know that submission starts a formal workflow or changes who can see a record.

Checkpoint placement must follow semantics rather than interface events. Asking for approval on every click defeats delegation. Asking immediately before a difficult-to-reverse submission concentrates attention where it matters. The checkpoint should show the proposed state change, affected records, policy basis, and uncertainty. A screenshot alone may not expose hidden consequences.

Validation is difficult because the same interface can display stale or adversarial content. A separate checker may need direct access to authoritative system state or a policy service. Comparing the final screen with the agent's goal is not enough when an indirect prompt injection changed the route or private data crossed an application boundary.

Recovery becomes distributed. One system may support undo while another has already sent a message or updated a downstream process. Contestability also becomes organizational. A coworker or customer affected by the change may know only that a record moved, not that an agent acted. A complete path therefore needs cross-system action identifiers, affected-party notice, ownership of correction, and evidence retention across application boundaries.

\subsection{Retail or airline service: policy-grounded action can still lack recourse}

Tau-bench models retail and airline service agents that interpret policy documents, interact with users, and call domain APIs \citep{yao2024bench}. ToolSandbox makes stateful execution and intermediate milestones visible \citep{lu2024toolsandbox}. ToolEmu uses an emulated sandbox to surface risks from high-stakes tool actions \citep{ruan2023identifying}. SABER distinguishes mutating actions and motivates verification before state change \citep{cuadron2025saber}. These artifacts make a booking or order path concrete.

Permission starts with the customer request. The request may be ambiguous about price, timing, travelers, refund conditions, or substitutions. A service policy constrains what the agent should do. The harness must also determine whether the customer has authority over every affected person and account.

The agent searches inventory, reads policy, proposes a selection, and calls an API. A trace can record each call. The decisive checkpoint belongs before purchase, cancellation, refund, or irreversible change. It should display the exact itinerary or order, final price, affected travelers or items, and applicable restrictions. A general confirmation such as ``shall I proceed?'' may be too compressed.

Independent validation can compare the proposed action against authoritative price, policy, identity, and inventory. It should not rely only on the natural-language summary produced by the acting model. Mutation-sensitive checks are promising because they target the transition where the external state changes. Their public description should still state what happens when the check fails and who can override it.

Recovery varies by domain. A mistaken cart update is easy to reverse. A nonrefundable ticket, canceled reservation, or issued refund may require compensation rather than rollback. The public artifact should identify these differences before the action is delegated.

Contestation closes the path. The customer needs an action identifier, the authority and policy applied, the current state, the provider responsible for review, and a remedy. Another traveler may need the same route even if they did not operate the agent. Without it, a technically well-logged booking remains unanswerable to the person who bears the cost.

\subsection{Cross-path result: the break moves, the structure persists}

The three paths differ in reversibility and affected parties. Repository work often has rich diffs and local rollback. Enterprise browser work crosses systems and roles. Service actions combine policy with money, identity, and other people's travel or orders. The location of risk changes.

The same structure persists. Permission is broader than the user's words, action is documented more clearly than the meaning of the state change, and traces are optimized for developers and evaluators. Checkpoints appear without timing; validators appear without visible independence. Recovery remains local while effects propagate, and contestability sits outside the runtime record.

This is why a feature checklist is insufficient. Every path can contain logging, human review, and error handling while still breaking between authorization and remedy. The action-path diagnostic asks where that break occurs and which audience loses the evidence needed to continue.

\section{An Action-Path Diagnostic}

The diagnostic is organized around continuity rather than certification. For one consequential action, a public artifact should allow a reader to answer seven questions.

\begin{enumerate}
\item \textbf{Permission:} Who delegated the action, what resources and people fall within scope, and when does that authority expire?
\item \textbf{Action:} Which tool call or state transition carried the consequence, and what data or account did it affect?
\item \textbf{Trace:} What evidence connects the request, policy, action, state change, and actor, and who can access that evidence?
\item \textbf{Checkpoint:} What condition pauses execution before the consequential change, and who can approve or reject it?
\item \textbf{Validation:} What checker uses which evidence, how is it independent from the actor, and can it stop the action?
\item \textbf{Recovery:} What can be undone, paused, compensated for, or repaired, by whom, and within what time?
\item \textbf{Contestation:} Who may challenge the action, how do they identify it, who reviews the challenge, and what remedy follows?
\end{enumerate}

The diagnostic should be applied at the action-class level before deployment and at the individual-action level after an event. A documentation page can state the general rules for sending messages, changing files, or making purchases. An action record can instantiate those rules with the specific permission, policy, trace, and remedy.

The seven stages should not become seven universal prompts. Low-stakes read-only actions may require minimal evidence and no human checkpoint. High-impact mutations need stronger checks. Risk sensitivity determines which path is appropriate. Public documentation should explain that mapping rather than advertise a generic human-in-the-loop feature.

\begin{table}[H]
\centering
\caption{Minimum public questions for an accountable action path.}
\begin{tabularx}{\textwidth}{p{0.17\textwidth}XX}
\toprule
Stage & Public question & Failure when missing \\
\midrule
Permission & Who authorized what scope? & Technical access substitutes for delegated authority. \\
Action & Which state change mattered? & Summaries hide the consequential transition. \\
Trace & What evidence connects request to effect? & Debugging records cannot support an account. \\
Checkpoint & Where can execution still be stopped? & Approval arrives after the decision that mattered. \\
Validation & What independent evidence checks the action? & The actor endorses its own reasoning. \\
Recovery & How can the effect be repaired? & Error documentation replaces remedy. \\
Contestation & How can an affected person obtain review? & Feedback exists without answerability. \\
\bottomrule
\end{tabularx}
\end{table}

\subsection{Diagnostic risk modes}

The seven-stage path locates continuity, while a complementary risk vocabulary names recurrent ways that continuity can fail. These are inspection categories grounded in the public corpus and adjacent literature. They are not observed incident frequencies, and one action may exhibit several at once.

\begin{table}[H]
\centering
\small
\caption{Diagnostic risk modes for action-path analysis. These categories support inspection and design; they are not prevalence estimates.}
\begin{tabularx}{\textwidth}{>{\raggedright\arraybackslash}p{0.20\textwidth}XX}
\toprule
Risk mode & Recognition rule & Runtime response to inspect \\
\midrule
Autonomy creep & Delegated scope expands through defaults, remembered permission, or reduced prompts. & Scoped identities, expiration, revocation, and least-privilege tools. \\
Misplaced checkpoints & Review occurs before the wrong step or after practical irreversibility. & Action-class gates before consequential state change. \\
Displaced verification labor & A stakeholder must judge the result after evidence has moved into inaccessible traces or summaries. & Role-specific evidence and review points. \\
Self-reinforcing validation & Actor and validator share a model, context, assumptions, or evidence source. & Independent evidence, stopping authority, and disclosed shared assumptions. \\
Delayed escalation & Handoff occurs after uncertainty or cumulative risk should have altered the path. & Escalation triggers tied to uncertainty, stakes, and downstream effects. \\
Poor recovery pathways & A bad action lacks undo, repair, compensation, containment, or redress. & Recovery designed by action class and consequence. \\
Overbroad logging & Evidence collection exceeds the accountability need or lacks access and retention limits. & Purpose limitation, minimization, role-based access, and bounded retention. \\
Third-party opacity & An affected non-user cannot identify, challenge, or correct the action. & Action identifiers, notice, standing, review ownership, and remedy. \\
\bottomrule
\end{tabularx}
\end{table}

The vocabulary distinguishes mechanism from harm. AgentDojo makes indirect prompt injection and tool misuse executable in a dynamic test environment \citep{debenedetti2024agentdojo}. Task Shield focuses on whether an agent's tool use remains aligned with the delegated task under indirect prompt injection \citep{jia2025thetask}. Agent Security Bench formalizes attack and defense categories for LLM-based agents \citep{zhang2025agent}. The OWASP taxonomy broadens the security account to risks such as privilege abuse, memory poisoning, insecure inter-agent communication, and cascading failure \citep{project2025owasp}. Our risk modes ask an additional set of questions: who authorized the tool call, when could it be stopped, what evidence remains, who can inspect that evidence, and how can the effect be repaired? A tool-misuse event, for example, may combine autonomy creep, a misplaced checkpoint, delayed escalation, and poor recovery. Naming the links avoids turning a complex action path into a single generic safety label.

The categories also keep third-party effects inside the analysis. Generative-AI incident research shows that consequences often fall beyond the direct user \citep{Li_2025}. In delegated-action settings, coworkers, clients, passengers, service recipients, and other users may bear an effect without seeing the original request or operating the agent. The action record therefore needs to preserve who authorized, configured, executed, validated, and could repair the act. Otherwise, technical observability for the builder coexists with opacity for the person who needs an answer.

\section{Design and Reporting Implications}

\subsection{Before execution: scope authority and concentrate attention}

Documentation should distinguish read-only inspection, local reversible edits, external communication, financial transactions, identity-bearing actions, and difficult-to-reverse changes. Each class should state the required permission, risk signals, checkpoint, and recovery route. This is more useful than a global autonomy level because the same agent can perform both trivial and consequential actions.

Permissions should name subjects as well as resources. Access to a shared calendar does not imply authority to cancel another person's meeting. Access to a repository does not imply authority over secrets or deployment. Access to a customer account does not imply authority over every traveler. Public artifacts should describe how third-party effects narrow the path.

Checkpoint design should minimize both interruption and consent fatigue. The system can batch low-risk actions, surface uncertainty, and pause before the first consequential mutation. The prompt should show the exact effect and remaining alternatives. A vague confirmation transfers verification labor back to the user without giving them useful evidence.

\subsection{During execution: build role-specific evidence}

An action record should connect the initiating request, permission scope, relevant policy, tool call, data touched, timestamp, validation, and escalation. The record should be tamper-evident enough for audit and minimal enough for privacy. Retention and access should follow the needs of each role.

User-facing views should foreground consequential steps and current state. Developer views can include detailed trajectories and errors. Auditor views need policy versions, credential scopes, model and harness versions, checkpoints, and overrides. Affected-party views should reveal only the action involving that person, its claimed basis, responsible provider, and remedy.

Validator independence should be reported explicitly. An artifact should not use ``verified'' as a free-standing claim. It should identify the validator, evidence source, separation from the actor, and stopping rule. Where full independence is impossible, the artifact should state the shared assumptions and residual risk.

\subsection{After execution: plan recovery before it is needed}

Recovery is easiest to design before deployment. Each action class should identify whether effects are reversible, compensable, pausable, or irreversible. Repository snapshots, transaction cancellation windows, message retraction, access revocation, and human escalation solve different problems.

The system should also identify downstream effects. Reverting a local file does not retract a release. Canceling a tool call does not remove copied data. Undo should be described at the level of consequence, not only command execution.

When recovery is impossible, the public artifact should say so and strengthen the checkpoint. This creates a direct connection between reversibility and supervision. The less repairable the action, the more evidence and independent review are justified before execution.

\subsection{During dispute: attach contestation to the action record}

Contestability should begin with an action identifier. A user or affected party needs to locate the event without reproducing the entire internal trace. The organization should state who reviews the challenge, which evidence is retained, how the actor or operator responds, and what remedies are available.

The route must remain available to non-delegating parties. A customer should not need access to an employee's agent console to correct a record. A colleague should not need the original prompt to challenge a message sent in their name. Public contact points should route to the responsible action record rather than a generic feedback queue.

Contestability also needs time. Evidence should persist long enough for a person to discover the effect and respond. Retention should still be bounded to avoid indefinite surveillance. These goals can be reconciled through action-class retention rules, role-based access, and selective preservation when a dispute begins.

\section{Discussion}

\subsection{The harness is where accountability is won or lost}

Cobbe and colleagues argue that accountable automated decision-making requires the review of an entire sociotechnical process, not only an explanation of a model output \citep{Cobbe_2021}. Kroll similarly treats traceability as a connection between technical records and the governance purposes those records are meant to serve \citep{kroll2021outlining}. We carry that process view into systems whose value depends on continuing after direct inspection recedes. In this setting, review after the fact is not enough. Authority, evidence, and intervention must remain connected while the action can still be changed.

The empirical asymmetry in our corpus is the hinge of this argument. Public artifacts routinely make tools and traces concrete, but rarely make checkpoint placement, validator independence, recovery, or contestability equally reconstructable. This does not show that deployed systems have no hidden safeguards. It shows that the public account of agentic capability is more mature than the public account of who can stop, judge, or repair the resulting action.

That imbalance changes how agent risk should be located. A highly capable model in a weak harness can receive broad permissions, act from compressed evidence, and leave little room for repair. A less capable model in a stronger harness may be more accountable because its action classes are bounded, its checkpoints precede consequential changes, and its validation does not merely reproduce the actor's judgment. Model safety remains necessary, but it cannot answer a question about the runtime arrangement by itself. For delegated action, the harness is not a wrapper around the accountability problem. It is where much of that problem is decided.

The reduced-supervision paradox therefore concerns a redistribution of labor and responsibility. Users gain time by not watching every step. The organization gains control over the infrastructure that substitutes for their attention. When that infrastructure exposes only what the agent did, it returns verification to the user after the fact and often after the action can no longer be changed. The efficiency benefit remains private to the system operator; the burden of proving error becomes socialized across users and affected parties.

\subsection{Observability can become an accountability substitute}

Ezell and colleagues argue that agent incidents require activity logs, system documentation, system access, and information about the tools involved \citep{ezell2025incident}. Their framework makes a persuasive case for richer evidence after failure. Our findings identify a complementary danger. Because tools and traces are already the most visible functions in the corpus, ``more observability'' can appear to close an accountability gap even when no one gains authority to intervene or obtain repair.

Chappidi and colleagues show that accountability-oriented record-keeping can reconfigure work, intensify surveillance, and generate tension between internal and external accountability demands \citep{Chappidi_2025}. That argument prevents an easy conclusion from our audit. The answer to thin accountability closure is not maximal logging. An exhaustive trace can still be unintelligible to a user, inaccessible to an affected third party, or disconnected from the forum able to require a remedy. It can also collect more worker and user data than the dispute requires.

We therefore make a stronger distinction between a trace and accountability evidence. A trace records an event. Accountability evidence connects the authority, policy, consequential state change, and responsible actor to a procedure that can question the action and alter its consequences. What counts as sufficient evidence must be judged by the forum that will use it, not by the maximum telemetry a system can retain. The same underlying event should support purpose-limited views without making indiscriminate surveillance the price of answerability.

\subsection{Delegation reaches beyond the user}

Riedl and Desai argue that computer science treatments of AI agents under-theorize loyalty and relations with third parties, even though both are central to legal agency \citep{riedl2025aiagentsandthelaw}. Our corpus gives that conceptual gap a runtime form. Third-party effects were clearly visible in 30 artifacts, yet this visibility rarely produced notice, standing, an external action identifier, or a route to challenge the act. The party who grants permission and the party who needs a remedy are often not the same person.

Alfrink and colleagues frame contestability as responsiveness to human intervention across the system lifecycle \citep{Alfrink_2022}. Our result does not dispute that lifecycle view; it exposes a continuity problem inside it. A system can contain interactive control during use and a complaint channel after deployment while failing to connect either mechanism to the same consequential action. Contestability then exists as a feature family but not as a usable path.

Kawakami and colleagues report that responsible-AI artifacts face barriers when they attempt to advance the goals of legal and civil-society stakeholders \citep{kawakami2024responsible}. The action-path analysis helps explain how such a barrier can arise. Builder-facing evidence does not automatically travel to the person affected by an agent's act. Nor does delegator consent settle the legitimacy of an action that changes another person's world. Accountability for agentic AI must therefore include people who never saw the prompt, never operated the interface, and may discover the action only through its consequences.

\subsection{From institutional duties to runtime controls}

The remaining step is institutional. The European Union's AI Act separates duties such as technical documentation, record-keeping, and human oversight rather than treating transparency as a single control \citep{reference2024regulation}. The NIST AI Risk Management Framework treats risk management as an organizational and lifecycle process \citep{technology2023risk}. Its generative-AI profile further specifies risks and actions across design, deployment, and use \citep{nationalinstituteofstandardsandtechnology2024artificialintelligenceriskmanagementframework}. The Model AI Governance Framework for Agentic AI brings that lifecycle orientation directly to agentic systems \citep{authority2026model}. These instruments differ in legal force and scope, but they share a translation problem: institutional commitments matter only if they alter a concrete action path.

Recent runtime-governance proposals attempt that translation. Koch maps governance norms onto enforceable runtime guardrails \citep{koch2026from}. Besanson places governance constraints in agent architecture rather than leaving them entirely in external policy \citep{besanson2026sarc}. The action-path diagnostic does not claim that either proposal, or any legal instrument, mandates its seven stages. It uses the stages as a test of institutional follow-through. If an organization claims human oversight, the action should contain an intervention point before the relevant consequence. If it claims responsibility, the action record should identify the role that owns review. If it claims redress, a person should be able to reach a repair process before the evidence disappears.

This test also distributes responsibility across the value chain. A model provider may document safeguards while a deployer grants overbroad credentials. A harness developer may expose detailed traces while a service organization offers no complaint route tied to an action identifier. A policy team may require approval while the interface requests it after payment or message delivery. A tool provider may execute a change without retaining the identifier needed by an auditor. Governance reaches runtime only when these institutional claims change what a particular action may do, what evidence it leaves, who can stop it, and who must repair it.

The resulting claim is intentionally provocative: an agentic system can be highly observable and still be institutionally unaccountable. Public visibility is not itself the remedy. It is useful when it preserves a path from delegated authority to consequential action and from evidence to intervention, judgment, and repair. When that path breaks, transparency documents the loss of control without restoring it.

\subsection{Alternative explanations and scope}

Source genre can explain part of the pattern. Research papers foreground novel technical contributions, while contestability may live in terms, support processes, or organizational policy. The non-research split showed the same direction, but the sample remains small. Direct deployment studies may find stronger internal controls.

Vocabulary can also obscure mechanisms. Recovery may be called snapshot, revert, compensation, or reconciliation. Contestability may appear as appeal, dispute, correction, or grievance. The codebook and retrieval probe used broad term families, yet no lexicon captures every domain. Row-level evidence and conservative partial codes reduce overclaiming.

Finally, some action settings may not require every stage. A read-only scientific query may have little need for rollback. A file deletion does. The diagnostic is not a universal compliance list. It asks designers to justify which stages apply, which are combined, and how the omission changes accountability.

\subsection{Research agenda}

The audit identifies what public artifacts make inspectable; the next research problem is whether the visible controls work. Deployment studies should therefore distinguish three states that current reporting often collapses: a documented control, a control observed during action, and a control that changes an outcome. Researchers can select consequential action classes and follow their permission scope, checkpoint timing, validator evidence, recovery, and dispute handling under standardized tasks.

Ibrahim and colleagues call for interactive evaluations that capture harms emerging over time rather than in isolated model outputs \citep{Ibrahim_2025}. Action-path studies can extend that proposal from interaction sequences to state-changing acts. Controlled experiments can vary the timing and content of checkpoints, then measure interruption cost, comprehension, approval quality, error detection, and consent fatigue. Validator studies can vary the model, evidence source, policy access, and stopping authority to test which forms of independence reduce shared failure.

Alfrink and colleagues treat contestability as a lifecycle design property \citep{Alfrink_2022}. That position should be tested from the perspective of people who did not delegate the agent. Interviews and complaint audits can examine whether affected parties receive notice, understand an action summary, locate the responsible provider, and obtain correction. The key outcome is not whether a complaint channel exists, but whether a person can connect it to the action that affected them.

Chappidi and colleagues show that accountability records alter organizational work rather than merely documenting it \citep{Chappidi_2025}. Organizational studies should therefore trace how one incident moves among developers, operators, legal teams, customer support, and auditors. Such work can reveal where evidence is translated, restricted, lost, or stripped of decision-making force.

Benchmarks can also extend beyond task success. They can score whether an agent stays within authority, retains the evidence appropriate to each audience, pauses before a high-impact mutation, accepts independent correction, recovers from error, and supports contestation. These measures should complement capability benchmarks. Collapsing them into one safety score would reproduce the abstraction that the action path is designed to expose.

\section{Limitations}

The corpus is targeted and overrepresents research papers. Coding-agent and benchmark artifacts are prominent because they expose runtime mechanisms clearly. English-language technical sources and stable public pages shape the sample. The findings cannot estimate how often controls appear or work in deployed systems.

The unit is a public artifact, not a product or organization. Several artifacts may describe related systems, and one system may distribute its accountability controls across papers, documentation, terms, and private operations. Present means visible in the inspected artifact. Unclear means the artifact could not support a stronger inference.

One primary coder conducted the analysis. Boundary examples, evidence notes, source-type sensitivity, and a retrieval probe improve traceability but do not establish independent agreement. Partial and unclear remain judgment-dependent. Later work should use independent coding before adjudication and report field-level agreement.

The retrieval probe measured lexical recovery, not semantic adequacy. A document can mention rollback without providing usable recovery. It can implement recovery using vocabulary outside the lexicon. The probe supports only the direction of public-language availability.

The paper proposes controls without evaluating them. Checkpoints can become burdensome. Validators can fail together. Recovery may be impossible after external propagation. Contestability can exclude people without technical, linguistic, or institutional access. These limitations are reasons to test the diagnostic, not evidence against the underlying need for connected action paths.

\section{Ethical Considerations}

This study analyzes public artifacts and uses no human participants or private system data. Its recommendations can still affect how organizations evaluate and document agentic systems. A checklist can become formal compliance without changing who can intervene or obtain repair. A visible action record can become a substitute for responsibility if no forum can question the organization that produced it.

Richer traces also create ethical risks. They can expose user data, worker activity, proprietary systems, and model context. We therefore reject the assumption that maximal logging is always desirable. Accountability evidence should be purpose-limited, role-specific, retained for a justified period, and accessible to people who need to question an action.

Access is part of that ethical boundary. A correction route can exclude affected people when it assumes technical expertise, dominant-language fluency, an account with the provider, or access to the delegating user's interface. Public documentation should identify who has standing, what minimum evidence a challenge requires, which organization owns review, and what alternatives exist when the person cannot access the internal trace. These requirements keep contestability from becoming another builder-facing feature.

\section{Conclusion}

Agentic AI creates value by conserving human attention. It can continue while the user attends to something else. Our audit shows that the public infrastructure surrounding this delegation is unevenly developed: across 63 artifacts, tool mediation and traces were far easier to inspect than checkpoint placement, validator independence, recovery, or contestability. The result is bounded to public visibility. It does not establish that deployed systems lack undisclosed controls or that visible controls work.

The reduced-supervision paradox gives this asymmetry a sharper interpretation. When direct inspection recedes, verification moves into the harness. If that harness preserves only action and trace, observability can legitimate delegation while returning the burden of detecting and proving error to users after meaningful intervention has passed. Repository, enterprise-browser, and service paths break in different places, but each reveals the same possibility: a technically reconstructable action can remain institutionally unanswerable.

We contribute an empirical account of that public-visibility imbalance, a theory of displaced verification, and an action-path diagnostic. The diagnostic follows one delegated act from permission to contestation and asks whether it remains bounded, reconstructable, interruptible, independently checkable, repairable, and challengeable. It also asks who can use the evidence. The user, operator, auditor, and affected third party may require different views, but those views must remain connected to the same consequential action.

Accountability therefore does not require people to watch every step. It requires the capacity to intervene at the step that matters and to repair the consequence afterward. An agentic system is not accountable merely because its actions are logged. Without authority, independent judgment, and a route to remedy, it is only well documented.

\bibliographystyle{plainnat}
\bibliography{refs}

\end{document}